# A Migratory Near Memory Processing Architecture Applied to Big Data Problems

Ed Upchurch
California Institute of Technology
1200 E. California Blvd. MS 158-79
Pasadena, CA 91125 USA
+1 626 395 2523
etu@caltech.edu

## ABSTRACT

The servers produced by mainstream vendors are highly inefficient in processing Big Data queries. The reason for this stems from bottlenecks that are inherent to the fundamental architecture of these systems. Today's server blades contain multicore processor chips connected to dynamic random access memory (DRAM) cards and disk drives by an interconnect chipset on the motherboard. The multicore processor chips perform all the calculations while the DRAMs and disk drives store the data, but have no processing capability. Thus in order to perform a query in a Big Data problem, the interconnection chipset must transfer data back and forth between the memory and a small storage area inside a processor called a cache, as well as between the memory and disk drives. For many Big Data applications the cache is too small, however, and the server must continuously get data from the disk or put it back there. Migratory Near Memory Servers alleviate these bottlenecks faced by conventional servers in handling Big Data systems by placing ultra-lightweight processors into the memory system. These ultra-lightweight processors run very simple programs that operate directly on the relations, vertices, and edges of Big Data systems directly where the data lies, without having to move it back and forth between the memory, cache and the heavyweight multicore processors. This paper addresses the application of such an architecture to relational database SELECT and JOIN queries. Preliminary results indicate orders of magnitude speed ups.

## 1. INTRODUCTION

The servers produced by mainstream vendors are highly inefficient in processing Big Data queries. The reason for this stems from bottlenecks that are inherent to the fundamental architecture of these systems. Today's server blades contain multicore processor chips connected to dynamic random access memory (DRAM) cards and disk drives by an interconnect chipset on the motherboard. The multicore processor chips perform all the calculations while the dual in-line memory cards (DIMMs) and disk drives store the data, but have no processing capability. Thus in order to perform a query in a Big Data problem, the interconnection chipset must transfer data back and forth between the memory and a small storage area inside a processor called a *cache*, as well as between the memory and disk drives. A typical server blade has several terabytes (TB) of disk storage, but only a few 100 gigabytes (GB) of memory, and a few megabytes (MB) of cache in the processors. For applications where the processor performs a long series of calculations on a small subset of data before requiring fresh data, the processor will almost always find the data it needs in the cache, and hence the conventional server organization works very well. This is definitely **not** the case in Big Data systems, however, and the server must continuously get data from the disk or put it back there. Because the time to write or read data to or from the disk is tens of thousands of times slower than it is to write or read memory in a DIMM, this is extremely costly and incredibly inefficient.

One way to improve the performance is to increase the amount of DRAM memory. Some high-end server blades can hold several terabytes of DRAM, but another bottleneck eventually limits Big Data application performance here as well. Although increasing the amount of memory reduces the frequency of access to disk, the relatively small number of cores on the blade, combined with the manner in which a large number of memory cards (DIMMs) are connected to the motherboard, limits the number of memory accesses that can be performed concurrently, a barrier known as the memory wall. Thus, even if a Big Data query could in theory be sped up by operating on different parts of a table or graph in parallel, the hardware has limited capability to support this. A new architecture designed for Big Data application requirements is needed.

## 2. NEW ARCHITECTURE: MIGRATORY NEAR MEMORY SERVER (MNMS)

Enhanced Memory Server blades may work alongside conventional server blades in the same rack to handle the data-intensive portions of Big Data applications far more efficiently than a rack filled with conventional blades alone [1]. The conventional servers do the jobs they do well such as inherently serial code, compute intensive code, housekeeping, operating systems functions and user network interfaces. Multiple racks are configurable into Big Data servers with a single, huge shared multi-petabyte memory capable of processing unprecedented numbers of concurrent transactions across this memory.

MNMS alleviates the bottlenecks faced by conventional servers in handling Big Data systems by placing ultra-lightweight processors—far simpler than the heavyweight (Intel, PowerPC, etc.) multicore processors on the motherboard—*into* the memory system [2]. These ultra-lightweight processors MNMS cores, run very simple programs called *threadlets* [3] that operate directly on the records, vertices, and edges of Big Data systems directly the data lies, without having to drag it back and forth between the memory, cache and the heavyweight multicore processors. Further, as threadlets scan records in Big Data applications or traverse the edges of Big Graph systems, they can *move* to the MNMS core closest to the data and resume execution. Finally, when data of particular interest is found they can "spawn" other threadlets to continue the processing when they move on. This unprecedented capability is not available in *any* other architecture. MNMS advantages are summarized in Table 1.

**Table 1. MNMS Advantages**

| | |
|---|---|
| **Low latency** | MNMS cores process data where it lies. |
| **High concurrency** | Large number of MNMS cores are distributed through the memory, each running a large number of threadlets against separate pieces of data |
| **High data bandwidth** | Because MNMS cores are connected directly to individual memory modules, they don't need to share wires to gain access to computation. MNMS core design is optimized to access data in the memory modules at the fastest rate possible supported by the memory. |
| **High memory density** | Because they don't need to support large heavyweight multicore processors, MNMS blades can pack more memory in each rack-mountable unit than conventional server blades. |
| **Low power overhead** | Only need is to supply power for the memory and ultra-lightweight MNMS cores, not for a heavyweight multicore processor. |
| **Low software overhead** | Threadlets logically see all the memory in a rack as a unified whole, called a *Partitioned Global Address Space* (PGAS) that eliminates the need for high overhead, inefficient layers of software to manage communications between blades. |

This paper describes results from considering SQL queries on a relational database. While there are several data base query languages being developed for Big Data applications such as NOSQL, SPARQL, RDF, The traditional SQL query language SELECT and JOIN were chosen for a case study of the MNMS architecture compared with traditional computer systems.

# 3. SELECT

A SELECT is the determination of which rows (tuples) of a table (relation) have some attributes that pass some tests (Boolean clause) often a case where looking for when attribute = select value. One or more attributes are identified and values to be compared against them provided. The simplest performance metrics for a SELECT are the response time and memory traffic (bytes that needed to be moved within the memory system) to find the next match.

Initially SELECT was explored on a generic MNMS topology for a variety of parameters, assuming that the entire relation is stored within MNMS memory (i.e. no disk accesses needed) and that the rows of the relation were scattered randomly throughout memory such that the entire relation had to be traversed in order to complete the SELECT (a kind of "worst case" scenario). The results below indicate a very substantial advantage for MNMS in terms of bandwidth as a function of total size of the attributes needed to define the SELECT. Gains of up to three orders of magnitude were possible in correspondence to a conventional architecture.

For classical cache-based architectures, the key parameters are the size of the cache line (especially relative to the size of a row entry and/or the size of an index entry), and how one gets from one entry to another.

- For selects on attributes that are not indexed, each row investigated requires at least one cache block to be dragged back to memory on a round trip pass. If the attribute being compared is bigger than a cache block, then additional block transfers will be necessary to get the whole attribute.
- If the attribute being selected on has an index, then the actual number of these round trip row reads that must be done goes down by something akin to the number of attribute/pointer pairs you can get in a cache block.

For MNMS generic architectures the key parameter is the size of the attribute that SELECT is operating on which is usually significantly less than the row size (at least an order of magnitude on average).

## 3.1 SELECT Results

Although many parameter sweeps such as database size, attribute size, and average number of responses were made, results are shown for the following case:

The scenario consists of a SELECT on a relation occupying a terabyte of memory having 31,250,000 rows (tuples). The classical system consists of a single host processor attached to a terabyte of RAM. The MNMS machine is made by replacing the terabyte of host RAM with a terabyte of MNMS memory nodes consisting of 8,000 MNMS cores (computed from the size of MNMS memory chips and number of cores/chip).

Sensitivity results are shown in Figure 1 for attribute size and average number of responses (as percentage of relation size). Results are shown for an average number of responses of 5% which is quite large for the relation size. The attribute size is varied from 8 bytes to 1000 bytes. Total data traffic in Mbytes is compared for the classical system and MNMS. Of course the traffic for the classical system is between RAM and the host (a long energy distance) while the MNMS traffic is within the memory nodes (a significantly shorter energy distance). MNMS delivers an estimated response time for the SELECT of 0.04 ms vs. 3125 ms for the classical case or a factor of 78,125 speedup for response time. This is probably the best comparison based on the difference in architectures – ie. simply replacing classical RAM with MNMS memory and doing away with the server.

Several observations were made from this initial model:

- MNMS's most sensitive parameter is the average number of responses
- The classical machine is not sensitive to the average number of responses as it has to read the entire relation in any case in this model.
- Both are mildly sensitive to attribute size with greatest sensitivity for very small attributes. The classical systems sensitivity is related to attribute size/cache line size.

In a more common case where the selectivity is much less than 1% MNMS, moves 100 to 1000 times less data.

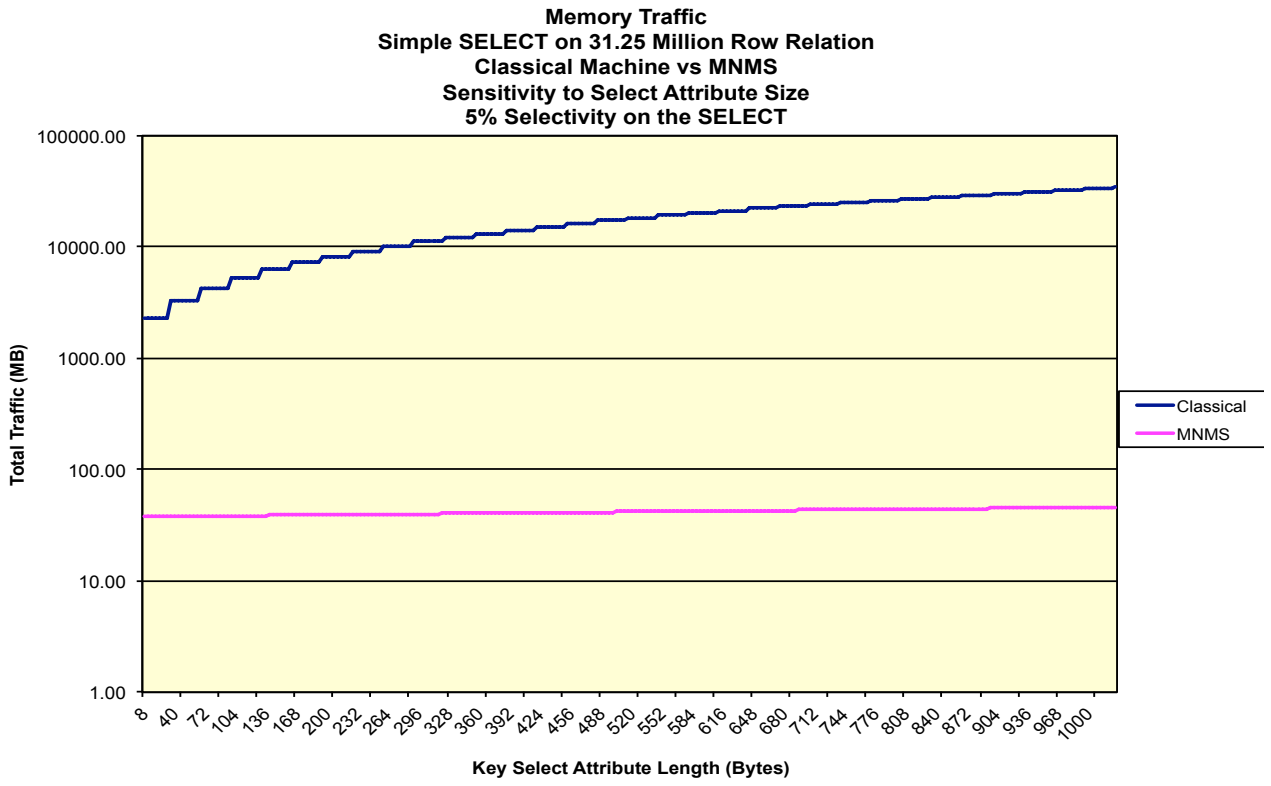

# 4. JOIN

Joins represent finding all pairs of rows from two tables where certain attributes match in some way. Join results consist of n tuples where each tuple of relation R is joined with exactly one tuple of relation S. Joins are used to compare or combine relations (rows) from two or more tables, when the relations share a common attribute value.

A simple algorithmic approach is: for every relation in the first table "S", loop over all relations in the other table "R", and see if the attributes match. This nested-loop algorithm is Order($n^2$) and one can usually do better. N-way joins are evaluated as a series of 2-way joins. For this project equijoins (find the tuples of R and S where a values of specific attributes of R and S and equal) were considered and hash algorithms. Initially a simple hash join with no indexing, no pre-hashing of all attributes for which joins are expected and no B-tree organization of the hashed results was considered.

In general a sequential hash join has two phases: (1) building where the inner relation is hashed into main memory using the join attributes as a hash key; (2) probing phase - the outer relation is read sequentially and for each record in the outer relation matching records of the inner relation are retrieved. Since each relation is read only once, 2n/cache line size reads are required for the classical machine. MNMS on the other hand looks at each tuple in the memory only once for each relation.

The expected speedups on a MNMS system are comparable to those for SELECT.

Classical algorithms for a join on a large database depend on whether or not the attributes being compared are indexed/sorted or not. If the attributes for both elements of the pairs are, then a merge join is all that is needed, which is a variation of a series of get next's. If either or both are not sorted or indexed, then either are (a) compared in all possible combinations, (b) sorted first (and at least one saved for reuse), and then a merge performed, or (c) have one side sorted and then do selects against the other side using in-order values from the first.

A parallel hash partitioned join splits the join into p smaller joins where p is the degree of parallelism based on the number of independent cores or MNMS memory units in the case of MNMS.

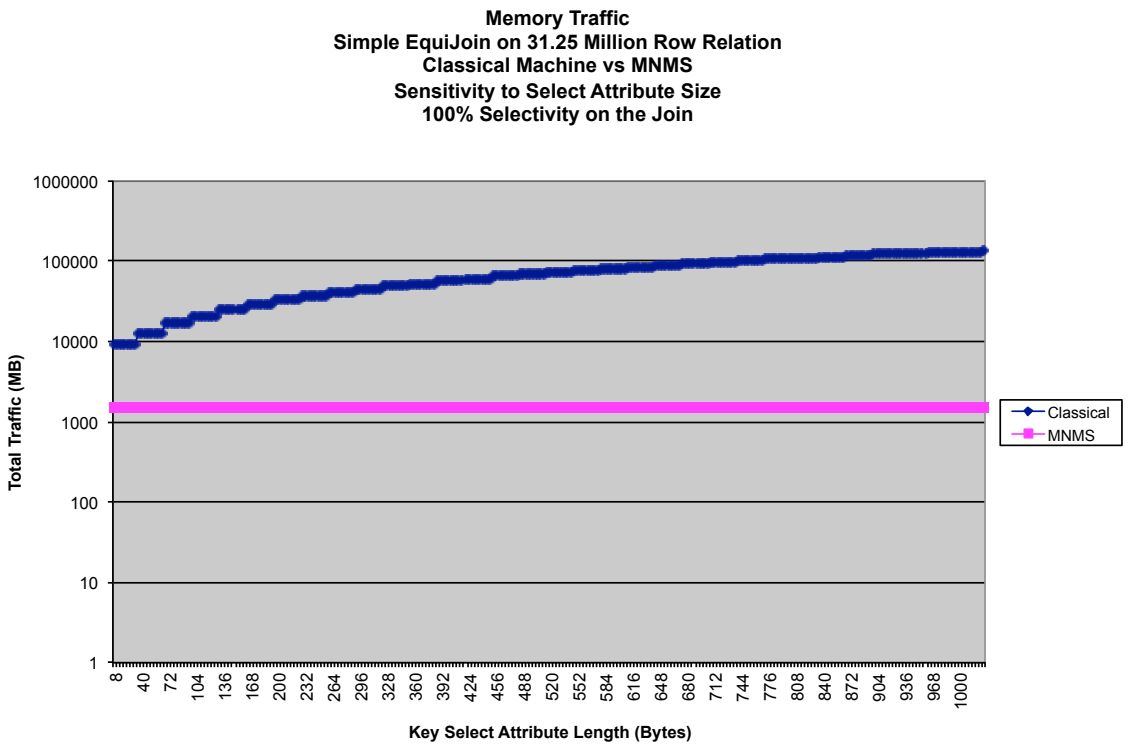

**Figure 2:** JOIN Memory Traffic

## 4.1 JOIN Results

JOIN attribute size, selectivity and complexity of JOIN (number of ways for multi-way JOIN) parameters were swept and data traffic compared with a conventional server model (same model as described for the SELECT above). The results of this model show that for a simple equijoin there is sensitivity to the JOIN attribute size as well as the JOIN selectivity (how many rows are returned by the JOIN that satisfy the JOIN attribute condition). Figure 2 shows the case for selectivity of 100% that is two equal sized relations (in this case 31,240,000 rows each) are joined creating another relation equal in size to the original relation size (31,250,000). This is not the worst case ($n^2$ rows would be that where n is the number of rows in each relation being joined) but it is very much on the high side of the result of an average join on most databases. With a selectivity of 100% one sees MNMS traffic being 1 to 2 orders of magnitude less than that of conventional servers. For 1% selectivity the difference is 3 to 4 orders of magnitude – difference is relatively linear in the selectivity for this simple model. As to attribute size two dynamics are going on: (1) for both architectures smaller join attribute sizes mean less data traffic as the request and response messages are smaller – the conventional server messages are always significantly larger than those of MNMS as they always get integral multiples of cache lines regardless of attribute size; (2) the traffic increase is proportional to (attribute size)/(row size). For the results shown the row size is 1000 bytes so as the JOIN attribute size approaches the row size the traffic differences due to attribute size is less and less for individual machines.

A more detailed JOIN model is being developed that includes the communication overhead of a fixed fanout generalized fat tree (FFGFT) topology [4]. In this model it is assumed that the MNMS memory compute/logical processing capability at each memory node is used to advantage to maintain a B-tree of attributes for each relation (or at least for each attribute class specified as "JOINable"). This tradeoff of space vs. traffic & response time yields an Order($\log_2 n$/[(#memory nodes)(#concurrent threads)]) JOIN algorithm – a JOIN can be done about as fast as a SELECT.

# 5. CONCLUSIONS

The MNMS architecture reduces memory traffic for relational SELECT and JOIN queries by several orders of magnitude for Big Data applications. This results in several orders of magnitude reducton in query response times and a significant reduction in energy consumption as less data is moved less distance than in conventional servers. The threadlet technology that moves processing to the data thus allows the efficient application of parallelism to the query processing.